
\magnification=1200
\voffset=0 true mm
\hoffset=0 true in
\hsize=6.5 true in
\vsize=8.5 true in
\normalbaselineskip=13pt
\def\doublespace{\baselineskip=20pt plus 3pt\message{double space}}
\def\singlespace{\baselineskip=13pt\message{single space}}
\let\spacing=\singlespace
\parindent=1.0 true cm



NUMBERING---------
\newcount\equationumber \newcount\sectionumber 
\sectionumber=1 \equationumber=1               
\def\setsection{\global\advance\sectionumber by1 \equationumber=1}
\def\numbe{{{\number\sectionumber}{.}\number\equationumber}
                            \global\advance\equationumber by1}
NUMBERING---
\def\numberit{\eqno{(\number\equationumber)} \global\advance\equationumber by1}
%
\def\numberal{(\number\equationumber)\global\advance\equationumber by1}
%
%
%
DOESN'T
%


\def\ccf#1{\,\vcenter{\normalbaselines
    \ialign{\hfil$##$\hfil&&$\>\hfil ##$\hfil\crcr
      \mathstrut\crcr\noalign{\kern-\baselineskip}
      #1\crcr\mathstrut\crcr\noalign{\kern-\baselineskip}}}\,}
\def\scf#1{\,\vcenter{\baselineskip=9pt
    \ialign{\hfil$##$\hfil&&$\>\hfil ##$\hfil\crcr
      \vphantom(\crcr\noalign{\kern-\baselineskip}
      #1\crcr\mathstrut\crcr\noalign{\kern-\baselineskip}}}\,}

\def\small3j#1#2#3#4#5#6{\def\st{\scriptstyle} 
   \bigl(\scf{\st#1&\st#2&\st#3\cr
           \st#4&\st#5&\st#6\cr} \bigr)}




\def\ref#1{$^{#1)}$}    


\def\upa#1{\raise 1pt\hbox{\sevenrm #1}}
\def\dna#1{\lower 1pt\hbox{\sevenrm #1}}
\def\dnb#1{\lower 2pt\hbox{$\scriptstyle #1$}}
\def\dnc#1{\lower 3pt\hbox{$\scriptstyle #1$}}
\def\upb#1{\raise 2pt\hbox{$\scriptstyle #1$}}
\def\upc#1{\raise 3pt\hbox{$\scriptstyle #1$}}
\def\hprime{\raise 2pt\hbox{$\scriptstyle \prime$}}
\def\ccom{\,\raise2pt\hbox{,}}


\def\asymptotically#1{\;\rlap{\lower 4pt\hbox to 2.0 true cm{
    \hfil\sevenrm  #1 \hfil}}
   \hbox{$\relbar\joinrel\relbar\joinrel\relbar\joinrel
     \relbar\joinrel\relbar\joinrel\longrightarrow\;$}}
\def\Asymptotically#1{\;\rlap{\lower 4pt\hbox to 3.0 true cm{
    \hfil\sevenrm  #1 \hfil}}
   \hbox{$\relbar\joinrel\relbar\joinrel\relbar\joinrel\relbar\joinrel
     \relbar\joinrel\relbar\joinrel\relbar\joinrel\relbar\joinrel
     \relbar\joinrel\relbar\joinrel\longrightarrow$\;}}

\catcode`@=11
\def\C@ncel#1#2{\ooalign{$\hfil#1\mkern2mu/\hfil$\crcr$#1#2$}}
\def\gf#1{\mathrel{\mathpalette\c@ncel#1}}      
\def\Gf#1{\mathrel{\mathpalette\C@ncel#1}}      

\def\gapx{\lower 2pt \hbox{$\buildrel>\over{\scriptstyle{\sim}}$}}
\def\lapx{\lower 2pt \hbox{$\buildrel<\over{\scriptstyle{\sim}}$}}

\def\nablaleft{\hbox{\raise 6pt\rlap{{\kern-1pt$\leftarrow$}}{$\nabla$}}}
\def\nablaright{\hbox{\raise 6pt\rlap{{\kern-1pt$\rightarrow$}}{$\nabla$}}}
\def\nablaboth{\hbox{\raise 6pt\rlap{{\kern-1pt$\leftrightarrow$}}{$\nabla$}}}

\def\boks#1#2{{\hsize=#1 true cm\parindent=0pt   
  {\vbox{\hrule height1pt \hbox
    {\vrule width1pt \kern3pt\raise 3pt\vbox{\kern3pt{#2}}\kern3pt
    \vrule width1pt}\hrule height1pt}}}}

\def\heading{ }
\def\range{ }

\def\body{\vfill\eject\parindent=1.0 true cm
 \ifx\spacing\singlespace\singlespace\else\doublespace\fi}
\def\title#1{\centerline{{\bf #1}}}

\def\today{\ifcase\month\or
  January\or February\or March\or April\or May\or June\or
  July\or August\or September\or October\or November\or December\fi
  \space\number\day, \number\year}
\let\date=\today
\newcount\hour \newcount\minute
\countdef\hour=56
\countdef\minute=57
\hour=\time
  \divide\hour by 60
  \minute=\time
  \count58=\hour
  \multiply\count58 by 60
  \advance\minute by -\count58

\def\sectionskip{\penalty-500\vskip24pt plus12pt minus6pt}

\def\sec{\hbox{\lower 1pt\rlap{{\sixrm S}}{\hbox{\raise 1pt\hbox{\sixrm S}}}}}
\def\section#1\par{\goodbreak\message{#1}
    \sectionskip\nobreak\noindent{\bf #1}\vskip0.3cm \noindent}

\nopagenumbers
\headline={\ifnum\pageno=\count31\frontheadline
  \else{\ifnum\pageno=0\frontheadline
     \else{{\raise 2pt\hbox to \hsize{\paperhead}}}\fi}\fi}

\footline={\centerline{\sevenbf \folio}}

\def\frontheadline{\sevenbf \hfil}
\def\paperhead{\sevenbf \heading\ \range\hfil\folio}
\newdimen\pagewidth \newdimen\pageheight \newdimen\ruleht
\maxdepth=2.2pt
\pagewidth=\hsize \pageheight=\vsize \ruleht=.5pt

\def\onepageout#1{\shipout\vbox{ 
    \offinterlineskip 
  \makeheadline
    \vbox to \pageheight{
         #1 
 \ifnum\pageno=\count31{\vskip 21pt\line{\the\footline}}\fi
 \ifvoid\footins\else 
 \vskip\skip\footins \kern-3pt
 \hrule height\ruleht width\pagewidth \kern-\ruleht \kern3pt
 \unvbox\footins\fi
 \boxmaxdepth=\maxdepth}
 \advancepageno}}

\output{\onepageout{\pagecontents}}

\count31=-1
\topskip 0.7 true cm

\pageno=0
\doublespace
\centerline{\bf Predictability in Quantum Gravity and Black Hole Evaporation}
\centerline{\bf J. W. Moffat}
\centerline{\bf Department of Physics}
\centerline{\bf University of Toronto}
\centerline{\bf Toronto, Ontario M5S1A7, Canada}
\vskip 2.0 true in
\centerline{\bf December, 1993}
\vskip 3 true in
{\bf UTPT-93-31}
\vskip 0.3 true in
{\bf e-mail: moffat@medb.physics.utoronto.ca}
\par\vfil\eject
\centerline{\bf Predictability in Quantum Gravity and Black Hole Evaporation}
\centerline{\bf J. W. Moffat}
\centerline{\bf Department of Physics}
\centerline{\bf University of Toronto}
\centerline{\bf Toronto, Ontario M5S 1A7, Canada}
\vskip 0.5 true in
\centerline{\bf Abstract}
\vskip 0.2 true in
A possible resolution of the information loss paradox for black holes is
proposed in which a phase transition occurs when the temperature of an
evaporating black hole equals a critical value, $T_c$, and Lorentz
invariance and diffeomorphism invariance are spontaneously broken.
This allows a generalization of Schr\"odinger's equation for the
quantum mechanical density matrix, such that a pure state can
evolve into a mixed state, because in the symmetry broken phase the
conservation of energy-momentum is spontaneously violated. TCP invariance
is also spontaneously broken together with time reversal invariance, allowing
the existence of white holes, which are black holes moving backwards in time.
Domain walls would form which separate the black holes and white holes
(anti-black holes) in the broken symmetry regime, and the system could evolve
into equilibrium producing a balance of information loss and gain.
\par\vfil\eject
{\bf Introduction}
\vskip 0.2 true in
It has been stated by Hawking$^{1}$ that quantum gravity introduces a new
level of unpredictability into physics over and above that associated with
the uncertainty principle, due to the emission of purely thermal radiation
by black holes. Ignoring problems with the semi-classical methods
of calculating the Hawking radiation, if the black hole disappears completely,
thereby
removing the information about the black hole states, then an initially pure
quantum state evolves into a mixed state. This is not permitted in standard
quantum mechanics, for a mixed state does not allow a precise determination
of any observable.

This situation, in which mixed states are produced by black hole evaporation,
arises when the system can be divided into two sections which do not interact
with each other. The Hilbert space ${\cal H}$ of the system is the tensor
product ${\cal H}_1\otimes{\cal H}_2$ of the Hilbert spaces of parts 1 and 2.
The Hilbert space ${\cal H}_1$ is defined in terms of particle states
at infinity and the Hilbert space ${\cal H}_2$ represents the states inside
the black hole.
The density operator $\rho$ of the total system is initially defined to
be in a
pure state. Suppose now that an observer can measure only part 1 of the system
and that he has no knowledge of part 2. Then he would assign equal
probability for all possibilities for part 2, and he would employ a
reduced density matrix $\tilde \rho$. In general, $\tilde \rho$ will describe
a mixed state on ${\cal H}_1$ even though $\rho$ describes a pure state
on ${\cal H}$.

If we acknowledge that quantum gravity can only be described by a mixed state,
composed of various pure quantum states, due to the loss of information in
black hole evaporation, then we would give up the notion that even in principle
one could work always with pure quantum states. This would, as Hawking has
proposed, would produce a degree of unpredictability into quantum gravity that
would profoundly change our view of the Universe.

Hawking replaced the standard Schr\"odinger equation for the density matrix
$\rho$ by a new evolution equation, which is still linear, and
first order in the time derivative:
$$
\dot \rho_{ab}={\tilde H}_{ab}^{cd}\rho_{dc},
\numberit
$$
where $\dot \rho=\partial/\partial t$ and the generalized Hamiltonian
operator $\tilde H$ must be constrained
to preserve the Hermiticity, positivity and trace of $\rho$. As shown
by Banks, Peskin and Susskind$^{2}$ (BPS), the most general equation preserving
$\rho^{\dagger}=\rho$ and $\hbox{Tr}\,\rho=1$ is of the form:
$$
\dot \rho=-i[H,\rho]-{1\over 2}k_{\alpha\beta}(Q^\alpha Q^\beta\rho
+\rho Q^\alpha Q^\beta - 2Q^\beta\rho Q^\alpha),
\numberit
$$
where the $Q$'s are Hermitian operators not equal to the identity, and
$k_{\alpha\beta}$ is a Hermitian matrix of coupling constants. Moreover,
$H$ is the Hamiltonian that appears in the commutator term. To guarantee
that $\rho > 0$, the eigenvalues of $k_{\alpha\beta}$ are positive or zero.
It was shown by BPS that, if $k_{\alpha\beta}$ has non-negative eigenvalues
and is, in addition, real and symmetric, violations of causality would ensue,
if we demand conservation of energy. It has also been
shown by Srednicki$^{3}$ that even if these causality violations are ignored,
then violations of Lorentz invariance would also occur.
\vskip 0.2 true in
{\bf 2. Spontaneous Breaking of Lorentz and Diffeomorphism Invariance.}
\vskip 0.2 true in
A scenario for early Universe cosmology has been proposed$^{4}$, in which
a Higgs mechanism is introduced into Einstein gravity, which spontaneously
breaks local Lorentz invariance and diffeomorphism invariance.
In this scheme, we postulate the existence of four scalar fields, $\phi^a$,
defined by
$$
\phi^a=e^a_\mu\phi^\mu,\quad \phi^\mu=e^\mu_a\phi^a,
\numberit
$$
where $e^a_\mu$ is a vierbein defined in terms of the metric:
$$
g_{\mu\nu}(x)=e^a_{\mu}(x)e^b_{\nu}(x)\eta_{ab}.
\numberit
$$
The vierbeins $e^a_{\mu}$ satisfy the orthogonality relations:
$$
e^a_{\mu}e_b^{\mu}=\delta^a_b,\quad e^{\mu}_ae_{\nu}^a=\delta^{\mu}_{\nu},
\numberit
$$
which allow us to pass from the flat tangent space coordinates (the fibre
bundle tangent space) labeled by $a,b,c...
$ to the world spacetime coordinates (manifold) labeled by $\mu,\nu,
\rho...$.

The fundamental form (4) is invariant under Lorentz transformations:
$$
e^{\prime\,a}_{\mu}(x)=L^a_b(x)e^b_{\mu}(x),
\numberit
$$
where $L^a_b(x)$ are the homogeneous $SO(3,1)$ Lorentz transformation
coefficients that can depend on position in spacetime, and which satisfy
$$
L_{ac}(x)L^a_d(x)=\eta_{cd}.
\numberit
$$

Let us assume that the
vacuum expectation value (vev) of the scalar fields, $<\phi^a>_0$,
will vanish for some temperature T less than a critical temperature $T_c$,
above
which the local Lorentz symmetry is broken. Above $T_c$ the non-zero
vev will break the symmetry of the gound state from $SO(3,1)$
down to $O(3)$. The four real scalar fields $\phi^a(x)$ are invariant
under Lorentz transformations:
$$
\phi^{\prime\,a}(x)=L^a_b(x)\phi^b(x).
\numberit
$$
The covariant derivative operator acting on $\phi^a$ is defined by
$$
D_{\mu}\phi^a=[\partial_{\mu}\delta^a_b+(\Omega_{\mu})^a_b]\phi^b,
\numberit
$$
where $(\Omega_\mu)^a_b$ denotes the spin connection.

We now introduce a Higgs sector into the Lagrangian density such
that the gravitational vacuum symmetry, which we set equal to the Lagrangian
symmetry at low temperatures, will break to a smaller symmetry at high
temperature. The pattern of vacuum phase transition that emerges contains
a symmetry anti-restoration$^{5}$. This vacuum symmetry breaking leads
to the interesting possibility that exact zero temperature conservation laws
e.g. electric charge and baryon number are broken in the early
Universe. In our case, we shall find that the spontaneous breaking of the
Lorentz symmetry of the vacuum leads to a spontaneous violation of the exact,
zero temperature conservation of energy.

Let us consider the Lorentz invariant Higgs potential:
$$
V(\phi^T s\phi )=-{1\over 2}\mu^2\phi^T s\phi+\lambda(\phi^T s\phi)^2,
\numberit
$$
where we choose $\lambda > 0$, so that the potential is bounded from below.
Here, we have introduced matrix notation and $s$ is an internal, field
dependent metric tensor $s_{ab}$, associated with the flat tangent space, which
is symmetric and positive and transforms as:
$$
s^\prime(x)=({L^{-1})}^T(x)s(x)L^{-1}(x).
\numberit
$$
Evidently, the form $\phi^T s\phi$ is invariant under local Lorentz
transformations.

Our Lagrangian density now takes the form:
$$
{\cal L}={\cal L}_G +\sqrt{-g}\biggl[{2\over f^2}Tr(D_\mu s s^{-1}D^\mu
ss^{-1})
+{1\over 2}D_{\mu}\phi^T sD^{\mu}\phi-{1\over 2}V(\phi^T s\phi)\biggr],
\numberit
$$
where ${\cal L}_G$ is the Lagrangian density for Einstein gravity:
$$
{\cal L}_G=-{c^4\over 16\pi G}\sqrt{-g}R,
\numberit
$$
$R=e^{\mu a}e^{\nu b}(R_{\mu\nu})_{ab}$ is the scalar curvature, and $f^2$
is a coupling constant.
The Lagrangian density (12) is invariant under Lorentz and diffeomorphism
transformations.

A calculation of the effective potential for the Higgs breaking contribution in
(12) shows that extra temperature dependent minima in the potential
$V(\phi)$ can occur for a non-compact group such as $SO(3,1)$$^{6}$, and
the spontaneous breakdown of $SO(3,1)$ to $SO(3)$ at high temperatures can
be realized.

We have introduced
the internal metric $s$, so as to guarantee unitarity of the $\phi$ matter
sector, which is then non-linearly realized
on the non-compact group $SO(3,1)$ and linearly realized on the maximal
compact subgroup $SO(3)$$^{7}$. The physical vacuum is not broken by this
spontaneous symmetry breaking when $<s>_0=s_0$, since
$<e_\mu>_0=<\Omega_\mu>_0
=0$ and $<\phi^T s\phi>_0=0$. The additional degrees of freedom associated
with $s$ are frozen out at low energies in that the coupling constant $f^2$
is chosen so that the internal metric terms in (12) only contribute
at energies of the order of the Planck mass.

If $V$ has a minimum at $\phi^T s\phi=v^2$, then the spontaneously broken
solution is given by $v^2=\mu^2/4\lambda$. There are three zero-mass
Goldstone bosons and after the spontaneous breaking of the vacuum one massive
physical boson particle $h$ remains.  A symmetry breaking
term occurs in the Lagrangian density:
$$
{\cal L}_{\Omega}={1\over 2}\sqrt{-g}(\Omega_{\mu})^{ab}v_b
(\Omega^{\mu})_a^cv_c
={1\over 2}\sqrt{-g}\sum_{i=1}^3((\Omega_{\mu})^{i0})^2(\mu^2/4\lambda).
\numberit
$$

A phase transition is assumed to occur at the critical temperature $T_c$,
when $v\not= 0$ and the Lorentz symmetry is broken and the three spin
connection
fields $(\Omega_{\mu})^{i0}$ and the associated Lorentz boost generators
are broken. Below $T_c$, the Lorentz symmetry is restored, and we
regain the usual classical gravitational field with massless graviton fields.

The total action for the theory is
$$
S_T=S+S_M,
\numberit
$$
where $S_T$ is given by
$$
S_T=\int d^4x {\cal L}_T,
\numberit
$$
and $S_M$ is the usual matter action for gravity.
Performing a variation of $S_T$ leads to the field equations:
$$
G^{\mu\nu}\equiv R^{\mu\nu}-{1\over 2}g^{\mu\nu}R={8\pi G\over c^4}(T^{\mu\nu}
+C^{\mu\nu}),
\numberit
$$
where $T^{\mu\nu}$ is the energy-momentum tensor for matter and $C^{\mu\nu}$
is the energy-momentum tensor for the matter field $\phi$ and the internal
metric field $s$.

Because $G^{\mu\nu}$ satisfies the Bianchi identities ${G^{\mu\nu}}_{;\nu}=0$,
we find in the broken symmetry phase:
$$
{T^{\mu\nu}}_{;\nu}=K^\mu,
\numberit
$$
where ; denotes the covariant derivative with respect to the Christoffel
symbols $\Gamma^{\lambda}_{\mu\nu}$, and $K^\mu$ contains terms
proportional to $v^2=<\phi^T s\phi>_0$.
Thus, the conservation of energy-momentum is spontaneously violated.
When the temperature passes below the critical temperature,
$T_c$, then $v=0$ and the
action is restored to its classical form with a symmetric degenerate
vacuum, and we regain the usual energy-momentum conservation laws:
$$
{T^{\mu\nu}}_{;\nu}=0,\quad {C^{\mu\nu}}_{;\nu}=0.
\numberit
$$
Let us perform a Lorentz transformation on $\phi^a$, so that we obtain$^{4}$:
$$
\phi^0=h,\quad \phi^1=\phi^2=\phi^3=0.
\numberit
$$
In this special coordinate frame, the remaining component $h$ is the physical
Higgs particle that survives after the Goldstone modes have been removed.
In this ``unitary gauge'' frame, we have
$$
S_h=\int d^4x\sqrt{-g}[{1\over 2}\partial_{\mu}h t
\partial^{\mu}h - V(h t h)],
\numberit
$$
where $t=s_{00}$ denotes the `time-time' components of the internal metric $s$.
The Hamiltonian associated with $S_h$ is bounded from below, i.e., there are no
ghost particles or tachyons in the particle spectrum.

 From the definition: $\phi^a=e^a_\mu\phi^\mu$, we get in the unitary
Lorentz frame, defined by (20):
$$
e^i_\mu=0,\quad i=1,2,3.
\numberit
$$
This produces a triangulation of the coefficients in general coordinate
transformations, causing a spontaneous breaking of diffeomorphism
invariance. This is an alternative proof that spontaneous violation of the
conservation of energy-momentum occurs in the broken symmetry vacuum.
\vskip 0.2 true in
3.{\bf Black Hole Evaporation and a Resolution of the Information Loss Problem}
\vskip 0.2 true in
The emission of Hawking radiation from a Schwarzschild black hole of mass
$M$ is identical in all respects to thermal emission from a perfect
black body at temperature T above absolute zero$^{8}$:
$$
T={\hbar c^3\over 8\pi kGM},
\numberit
$$
where $k$ is Boltzmann's constant. As the evaporation proceeds, the mass
$M$ decreases and the temperature $T$ increases and eventually
$T$ will equal the critical
temperature $T_c$, at which a first order phase transition occurs that
spontaneously breaks local Lorentz symmetry. Since in the broken symmetry
phase of the black hole evaporation the conservation of energy-momentum
is spontaneously broken, we are able to provide a physical picture for
the generalized
Schr\"odinger equation for the density matrix $\rho$ given by (2). Thus, we
can now have a pure quantum state decay into a mixed state
during the broken symmetry phase of the evaporation of the black hole.
Once the black hole completely evaporates, then the spacetime symmetries
and the conservation of energy are restored.

There are two other possible scenarios besides the one contemplated
above$^{1,9}$:
\vskip 0.1 true in
1. A naked singularity of negative mass is produced as a final product of the
evaporation.
\vskip 0.1 true in
2. A remnant black hole of about the Planck mass is left after some
mechanism stops the evaporation.
\vskip 0.1 true in
Possibility 1 would result in a complete breakdown of predictability due to
a large
number of negative mass naked singularities being formed in the early Universe.
Also, as would apply to possibility 2, the density of the Universe would
be dominated by naked singularities and black hole remnants, which would
give rise to unreasonable values for the critical density parameter
$\Omega$ and the deceleration parameter.
However, we must allow the possibility that a future solution to
quantum gravity will result in the removal of singularities as a final product
of black hole evaporation, in which case scenario 1. would no longer
be viable.

Scenario 2 suffers from the problem that a large amount of information
about the black hole states would have to be emitted in the final stages of
evaporation. The time during which the information was emitted would have
to be unreasonably long, $\sim \hbox{exp}(4\pi M^2/M_p^2)/M_p$, where $M_p$
is the Planck mass. This would cause similar problems with the sizes of
$\Omega$ and the deceleration parameter.

Thus, the scenario in which the black hole disappears completely, erasing
any information about the black hole states and any conserved quantities
that are coupled to long range fields, is the most reasonable one. The
spontaneous violations of Lorentz invariance and conservation of
energy-momentum in the symmetry broken phase of evaporation, at temperatures
of order the Planck temperature, would permit a stage in which the initial pure
quantum state of the system will have decayed into a mixed state
in which one cannot make any exact predictions about the outcome of
experiments, but only probabilities for the different possible outcomes.

Hawking$^{1}$ modified his quantum mechanics so that it was possible to have
a pure state decay into a mixed state globally. The intention was that
any serious violation of known physical laws would not be detected by
an observer at infinity, i.e., somehow a violation of probabilities could be
confined to a small region of spacetime. But BPS showed that this was not the
case. Conservation of energy would be accompanied by large violations of
causality at macroscopic levels, vitiating the physical scenario. In contrast,
our model confines the modification of quantum mechanics to a region of the
order of the Planck volume after the onset of the phase transition
for $T\sim T_c$. Quantum mechanics remains unaltered in the initial phase
of evaporation, since the broken symmetry phase is hidden by the event
horizon. At the phase transition, the symmetry broken phase
and the modified quantum mechanical regime will be revealed to an external
observer, at which time he will become aware that information
was being lost through the violation of energy and entropy
conservation inside a region within the black hole. Thus, the observer
will then ``know'' why he has witnessed the decay of the pure quantum
state-- describing the original collapsed star-- into a mixed state with
the accompanying loss of information. At the phase transition temperature,
$T_c$, the evaporating black hole will have the same mass, $M_c$, and radius,
$R_c$, and all black holes will evolve in an identical fashion as they shrink
to zero mass. The precise way in which the information is lost dynamically,
through the spontaneous violation of the conservation of energy and entropy,
is not understood at present, for a knowledge of this mechanism depends
upon being able to solve the field equations in the broken symmetry phase.
Although the loss of information (or the associated amount of gained entropy)
is subject to the same unpredictability as any other physical observable in
the modified quantum mechanical region, we must require a probabilistic
accounting of information loss or gain.

We must expect that the decay of a pure quantum state into a mixed state should
occur for microscopic black holes at the elementary particle level, for
quantum fluctuations of the metric can be interpreted as virtual black
holes which appear and disappear. The time reversal of a black hole spacetime
diagram is a white hole, which is analogous to an electron being the time
reversed state of a positron, i.e. a white hole is the ``anti-blackhole''
state and black holes and white holes will create and annihilate in the
vacuum. In the symmetry broken phase of these virtual black holes and
anti-black holes, which is confined to a localized region of spacetime,
the anti-black holes will {\it produce} information or, in other words,
decrease entropy leading to an overall energy balance in the vacuum.

Within the framework of a
Lorentz invariant, local field theory, assuming the usual connection between
spin and statistics, invariance under time reversal is equivalent to
invariance under $P C$ i.e., the combined operation of charge conjugation
($C$) and space inversion ($P$). In the broken Lorentz symmetry phase,
the L\"uders, Pauli, Schwinger$^{10}$ TCP theorem is spontaneously broken,
and in our scheme there will be a {\it localized} violation of time reversal
invariance under the operation of time inversion $T$. Thus, in the broken
symmetry phase, the time arrow of the second law of thermodynamics can be
reversed, permitting the existence of white holes (or anti-black holes).

We can also postulate the existence of a macroscopic white hole in the final
symmetry broken phase of evaporation, since the arrow of time can be reversed
in this phase, producing this macroscopic white hole (or anti-black hole) and
the latter will produce information or, equivalently, entropy decrease. The
situation could then evolve to an information balance in the
Universe, before the white hole and black hole evaporate away. Due to the
spontaneous breaking of the discrete time reversal symmetry, associated with
the phase transition, a domain wall$^{11}$ will form that separates the white
hole from the black hole within the symmetry broken regime.
\par\vfil\eject
{\bf Acknowledgements}
\vskip 0.2 true in
I thank M. Clayton, N. Cornish and P. Savaria for stimulating discussions. This
work was supported by the Natural Sciences and Engineering Research Council of
Canada.
\vskip 0.2 true in
\centerline{\bf References}
\vskip 0.2 true in
\item{1.}{S. W. Hawking, Phys. Rev. D {\bf 14}, 2460 (1976);
Commun. Math. Phys. {\bf 87}, 395 (1982).}
\item{2.}{T. Banks, L. Susskind, and M. E. Peskin, Nucl. Phys. B{\bf 244},
125 (1984).}
\item{3.}{M. Srednicki, University of California preprint, hep-th/9206056
UCSBTH-92-22, revised July 1993.}
\item{4.}{J. W. Moffat, Found. of Phys. {\bf 23}, No. 3, 411 (1993);
Int. J. of Mod. Phys. D {\bf 2}, September issue (1993).}
\item{5.}{L. D. Landau and E. M. Lifshitz, {\it Statistical Physics},
translated by J. B. Sykes and M. J. Kearsley, Addison-Wesley Publishing
Company, Mass. p.427; S. Weinberg, Phys. Rev. D {\bf 9}, 3320 (1974);
R. Mohapatra and G. Senjanovic, Phys. Rev. D {\bf 20}, 3390 (1979);
P. Langacker and So-Young Pi, Phys. Rev. Letts. {\bf 45}, 1
(1980); V. Kuzmin, M. Shaposhnikov, and I. Tkachev, Nucl. Phys. {\bf B196},
29 (1982); T. W. Kephart, T. J. Weiler, and T. C. Yuan, Nucl. Phys.
{\bf B330}, 705 (1990); S. Dodelson and L. M. Widrow, Phys. Rev.
D {\bf 42}, 326 (1990).}
\item{6.}{P. Salomonson and B. K. Skagerstam, Phys. Lett. B{\bf 155},
98 (1985).}
\item{7.}{The use of an internal metric in this kind of theory was first
proposed by K. Cahill, Phys. Rev. D {\bf 18}, 2930 (1978); {\bf 20}, 2636
(1979);
J. Math. Phys. {\bf 21}, 2676 (1980); Phys. Rev. D {\bf 26}, 1916 (1982);
see also J. E. Kim and A. Zee, {\it ibid.} {\bf 21}, 1939 (1980);
B. Julia and F. Luciani, Phys. Lett. {\bf B90}, 270 (1980); J. Dell, J. L.
de Lyra and Lee Smolin, Phys. Rev. D {\bf 34}, 3012 (1986); D. S. Popovi\'c,
Phys. Rev. D {\bf 34}, 1764 (1986). For a discussion of unitarity problems
associated with non-compact groups, see: J. P. Hsu and M. D. Xin, Phys. Rev.
D {\bf 24}, 471 (1981) and ref. 4.}
\item{8.}{S. W. Hawking, Commun. Math. Phys. {\bf 43}, 199 (1975); J. D.
Bekenstein, Phys. Rev. D {\bf 7}, 2333 (1973);
R. M. Wald, Commun. Math. Phys. {\bf 45}, 9 (1975); Phys. Rev. D {\bf 9},
3292 (1974); L. Parker, Phys. Rev. D {\bf 12}, 1519 (1975).}
\item{9.}{For reviews, see: J. Preskill, ``Do Black Holes Destroy Information?"
California Institute of Technology preprint CALT-68-1819 (1992); S. B.
Giddings, Phys. Rev. D {\bf 46}, 1347 (1992); ``Toy Models for Black Hole
Evaporation" UCSBTH-92-36 hep-th/9209113 (1992).}
\item{10.}{J. Schwinger, Phys. Rev. {\bf 82}, 664 (1951); W. Pauli, {\it Niels
Bohr and the Development of Physics}, McGraw Hill, New York (1955);
G. L\"uders, Ann. of Phys. (New York), {\bf 2}, 1 (1957).}
\item{11.}{J. Preskill, Ann. Rev. Nucl. Part. Sci. {\bf 34}, 461 (1984);
A. Vilenkin, Phys. Rep. {\bf 121}, 263 (1985).}

\end